\documentclass{article}
\usepackage{spconf}
\usepackage{amsmath}
\usepackage{amssymb}
\usepackage{subfigure}
\usepackage{caption}
\usepackage{lineno}
\usepackage{graphicx}
\usepackage{epstopdf}
\usepackage{array}
\usepackage{setspace}
\title{Accelerated Magnetic Resonance Spectroscopy with Vandermonde Factorization}

\begin{document}
\name{Xiaobo Qu$^{\star 1}$ \qquad Jiaxi Ying$^{1}$  \qquad Jian-Feng Cai$^{2}$ \qquad Zhong Chen$^{1}$ \thanks{*Corresponding author: Xiaobo Qu (quxiaobo@xmu.edu.cn)}}
\address{$^{1}$ Department of Electronic Science, Xiamen University, Xiamen, China \\
$^{2}$ Department of Mathematics, Hong Kong University of Science and Technology, Hong Kong SAR, China}

\maketitle

\begin{abstract}
Multi-dimensional magnetic resonance spectroscopy is an important tool for studying molecular structures, interactions and dynamics in bio-engineering. The data acquisition time, however, is relatively long and non-uniform sampling can be applied to reduce this time. To obtain the full spectrum,a reconstruction method with Vandermonde factorization is proposed.This method explores the general signal property in magnetic resonance spectroscopy: Its time domain signal is approximated by a sum of a few exponentials. Results on synthetic and realistic data show that the new approach can achieve faithful spectrum reconstruction and outperforms state-of-the-art low rank Hankel matrix method.
\end{abstract}
\section{Introduction}
\label{sec:intro}

Magnetic resonance spectroscopy (MRS) has been regarded as an indispensable tool in studying a molecular structure and dynamics or interactions of biopolymers in chemistry and biology. However, the duration of a multi-dimensional MRS experiment is proportional to the number of measured data points and increases rapidly with spectral resolution and dimensionality. The non-uniform sampling (NUS) approach offers a general solution for a dramatic reduction in measurement time. Reconstructing the full signal from a non-uniformly sampled signal is essential for the next step of data analysis. The reconstruction may be successful by exploiting the inherent structure of the signal in the time or frequency domains.

One line of work is concerned with the sparsity of MRS in the frequency domain. Compressed sensing (CS) \cite{Candes2006} suggests that if the signal enjoys a sparse representation in some transform domain, it is possible to recover a signal even when the number of samples is far below its ambient dimension. CS has been demonstrated as an effective tool for reconstructing NUS spectrum by assuming that the spectrum is sparse in the frequency domain \cite{Qu-sparse-2010, Kazimierczuk2011, Qu2011}. However, broad peaks inevitably results in loss of sparsity since more non-zeros are presented in the spectrum \cite{Qu-Accelerate-2015}. This limitation of CS lead to the degeneration of reconstruction performance.

Another line of work is concerned with the low rankness of MRS signals in the time domain. The time domain signal of MRS is called the free induction decay (FID), which is generally approximated by a sum of a few decaying exponential functions. By transforming the FID into a so called Hankel matrix, the number of exponentionals will equal to the rank of this matrix \cite{Zhipei-2013}. Assuming the number of spectral peaks is much smaller than the FID data points, the low rank Hankel matrix completion (LRHMC) \cite{Qu-Accelerate-2015} was proposed to recover the FID signals in the NUS MRS. Unlike CS, which seeks the sparsity of the spectrum in the frequency domain and encounters problems to represent the signals with fast decay, LRHMC tried to minimize the number of peaks. Experimental results on simulated and real MRS data show that broad peaks can be recovered by LRHMC much better than the ${l_1}$ norm minimization on the spectrum \cite{Qu-Accelerate-2015}. The recovery condition of low rank Hankel matrix under uniformly random sampling and Gaussian random encoding can be found in \cite{Chen-Chi2014} and \cite{Cai-RobustLR-2016}, respectively.

In LRHMC, the connection between exponentials and matrix structures is still not fully explored. For example, it is still unclear how the signal subspace is related to each specific exponential function. Taking this into account, we propose to use the Vandermonde structure of the Hankel matrix to reconstruct the FID. These exponentials are explicitly presented in the Vandermonde decomposition of the Hankel matrix. This nice property allows enforcing more matrix structures in the signal model thus has great potential to improve the spectrum reconstruction. Experiment results on synthetic data and real MRS data show that the new approach requires significantly fewer measurements than LRHMC to achieve a faithful reconstruction of the FID.

The rest of this paper is organized as follows. Section \ref{sec:II} introduces related work. Section \ref{sec:III} presents the proposed  Hankel matrix completion with Vandermonde factorization (HVaF) method . Section \ref{sec:VI} presents the numerical results on both simulated and real-world data. Section \ref{sec:Discussion} concludes this work and discusses some future work.

\section{Related work}
\label{sec:II}

Let ${\cal R}$ be a Hankel operator which maps a vector ${\bf{x}} \in {\mathbb{C}^n}$ to a Hankel matrix ${\cal R}{\bf{x}} \in {\mathbb{C}^{{n_1} \times {n_2}}}$ with ${n_1} + {n_2} = n + 1$ as follows
\begin{equation*}
{[{\cal R}{\bf{x}}]_{ij}} = {x_{i + j}},{\kern 3pt} \forall i \in \{ 0, \ldots ,{n_1}{\rm{ - }}1\} ,{\kern 3pt}j \in \{ 0, \ldots ,{n_2} - 1\} ,
\end{equation*}
In particular, we denote the Hankel operator by ${\cal H}$ instead of ${\cal R}$ in the case ${n_1} = {n_2}$.

The LRHMC \cite{Qu-Accelerate-2015} is based on the observations that the Hankel matrix ${\cal R}{\bf x}$ constructed by the FID ${\bf{x}}$ is low rank if the number of spectral peaks is much smaller than the data points in the whole spectrum. Hence, the reconstruction problem can be formulated as the low rank matrix completion problem
\begin{equation}\label{eq:LRHMC}
\mathop {\min }\limits_{\mathbf{x}} {\kern 3pt} {\left\| {\mathcal{H}\mathbf{x}} \right\|_*} + \frac{\lambda }{2}\left\| {{\mathbf{y}} - {\mathbf{Dx}}} \right\|_2^2,
\end{equation}
where ${\bf x}$ is the acquired NUS FID data, ${\bf D}$ is an operator of the NUS schedule, ${\left\|  \cdot  \right\|_*}$ is the nuclear norm defined as a sum of matrix singular values. The efficiency of LRHMC has been verified on numerical simulations and real MRS data \cite{Qu-Accelerate-2015}. However, we will present in Section \ref{sec:III} that the Hankel matrix of the FID has Vandermonde factorization and experiment results in Section \ref{sec:VI} show that the new approach exploiting Vandermonde factorization can achieve much better reconstruction than LRHMC from the same NUS data.

\section{Hankel matrix completion with Vandermonde factorization}
\label{sec:III}
The general signal property in MRS that FID can be approximated by a sum of a few decaying exponentials has been widely acknowledged \cite{Qu-Accelerate-2015,Zhipei-2013}. Let vector ${\mathbf{y}} \in {\mathbb{C}^{2N - 1}}$ be the complete FID
\begin{equation}\label{eq:signal}
{y_k} = \sum\limits_{r = 1}^R {{d_r}{e^{ - k\Delta t/{\tau _r} + i2\pi {f_r}k\Delta t}}},
\end{equation}
where ${d_r}$, ${\tau _r}$ and ${f_r}$ are the complex amplitude, decay time and frequency, respectively, of the $k$-th exponential.

Let define ${z_r} = {e^{ - \Delta t/{\tau _r} + i2\pi {f_r}\Delta t}}$. It is observed that the Hankel matrix of the FID ${\mathbf{y}} \in {\mathbb{C}^{2N - 1}}$ admits a Vandermonde factorization
\begin{equation}\label{eq:VandF}
\begin{array}{l}
\begin{split}
{\cal H}{\bf{y}} = \left[ {\begin{array}{*{20}{c}}
1& \cdots &1\\
{{z_1}}& \cdots &{{z_R}}\\
 \vdots & \vdots & \vdots \\
{z_1^{N - 1}}& \cdots &{z_R^{N - 1}}
\end{array}} \right]\left[ {\begin{array}{*{20}{c}}
{{c_1}}&{}&{}\\
{}& \ddots &{}\\
{}&{}&{{c_R}}
\end{array}} \right]\\
\left[ {\begin{array}{*{20}{c}}
1&{{z_1}}& \cdots &{z_1^{N - 1}}\\
 \vdots & \vdots & \cdots & \vdots \\
1&{{z_R}}& \cdots &{z_R^{N - 1}}
\end{array}} \right].
\end{split}
\end{array}
\end{equation}
Obviously, \eqref{eq:VandF} can be easily rewritten as a form of the product of two factor matrices and this paper focus on the latter form. Note that we choose the dimension of ${\bf{y}}$ to be odd simply to yield a squared matrix ${\cal H}{\bf{y}}$. Actually, our algorithm and results do not rely on the dimension being odd.

In this paper, we formulate the signal reconstruction as Hankel matrix completion with Vandermonde factorization. Specifically, we aim to find ${\bf{x}}$, the Hankel matrix of which can be factorized into two Vandermonde matrices, i.e., $\mathcal{H}{\mathbf{x}} = {\mathbf{U}}{{\mathbf{V}}^H}$ with ${\bf U}$ and ${\bf V}$ Vandermonde matrices. To force ${\bf U}$ and ${\bf V}$ to be Vandermonde matrices is equal to restrict each column of ${\bf U}$ and ${\bf V}$ to contain single component of exponential function. According to Kronecker’s theorem \cite{Andersson-FreqEstim-2014}, we can equally constrain the Hankel matrix of each column of ${\bf U}$ and ${\bf V}$ to be rank 1. Therefore the optimization \eqref{eq:VandF} can be equivalently rewritten as
\begin{equation}\label{eq:RankVand}
{\rm{find}}{\kern 5pt} {\bf{x}},{\bf{U}},{\bf{V}}
\end{equation}
\[s.t.{\kern 5pt} {\text{rank}}(\mathcal{R}({{\mathbf{U}}_{(:,r)}})) = 1,{\kern 3pt} {\text{rank}}(\mathcal{R}({{\mathbf{V}}_{(:,r)}})) = 1, \]
\[\mathcal{H}{\mathbf{x}} = {\mathbf{U}}{{\mathbf{V}}^H},{\kern 3pt} {\mathbf{y}} = {\mathbf{Dx}},{\kern 3pt} {\text{for}}{\kern 3pt} {\text{all}}{\kern 3pt} r \in \{ 1, \ldots ,K\}.\]

Given that it is difficult to develop a reliable and computational algorithm to solve the rank-constrained problem and the measurements in MRS experiments are usually contaminated by bounded noise, we relax \eqref{eq:RankVand} and solve the following problem instead
\begin{equation}\label{eq:NuclearNormVand}
\mathop {\min }\limits_{{\mathbf{U}},{\mathbf{V}},{\mathbf{x}}} \sum\limits_{r = 1}^K {({{\left\| {\mathcal{R}{\mathcal{Q}_r}{\mathbf{U}}} \right\|}_*} + {{\left\| {\mathcal{R}{\mathcal{Q}_r}{\mathbf{V}}} \right\|}_*})}+ \frac{\lambda }{2}\left\| {{\mathbf{y}} - {\mathbf{Dx}}} \right\|_2^2.
\end{equation}
\[s.t. \mathcal{H}{\mathbf{x}} = {\mathbf{U}}{{\mathbf{V}}^H}.\]

To solve \eqref{eq:NuclearNormVand}, we develop an algorithm based on half quadratic methods with continuation for its advantage in handling multivariable optimization \cite{ Qu2014}. We introduce the term $\left\| {{\cal R}{\bf{x}} - {\bf{U}}{{\bf{V}}^H}} \right\|_F^2$ to keep ${\cal R}{\bf{x}}$ close enough to ${\bf{U}}{{\bf{V}}^H}$ instead of addressing the constraint ${\cal R}{\bf{x}} = {\bf{U}}{{\bf{V}}^H}$ directly, and propose the following optimization,
\begin{equation}\label{eq:HVaF}
\begin{split}
&\mathop {\min }\limits_{{\mathbf{U}},{\mathbf{V}},{\mathbf{x}}} \sum\limits_{r = 1}^K {({{\left\| {\mathcal{R}{\mathcal{Q}_r}{\mathbf{U}}} \right\|}_*} + {{\left\| {\mathcal{R}{\mathcal{Q}_r}{\mathbf{V}}} \right\|}_*})} + \frac{\beta }{2}\left\| {\mathcal{H}{\mathbf{x}} - {\mathbf{U}}{{\mathbf{V}}^H}} \right\|_F^2\\
&{\kern 50pt}+ \frac{\lambda }{2}\left\| {{\mathbf{y}} - {\mathbf{Dx}}} \right\|_2^2.
\end{split}
\end{equation}
When $\beta  \to \infty $, the solution to \eqref{eq:HVaF} is approaching to \eqref{eq:NuclearNormVand}. To solve \eqref{eq:HVaF}, some auxiliary variables are introduced into model \eqref{eq:HVaF}, and then \eqref{eq:HVaF} is reformulated into the following equivalent form:
\begin{equation}\label{eq:AuxiMode}
\begin{split}
&\mathop {\min }\limits_{{\bf{U}},{\bf{V}},{\bf{x}}} \sum\limits_{r = 1}^K {({{\left\| {{{\bf{B}}_r}} \right\|}_*} + {{\left\| {{{\bf{C}}_r}} \right\|}_*})}  + \frac{\beta }{2}\left\| {{\cal H}{\bf{x}} - {\bf{U}}{{\bf{V}}^H}} \right\|_F^2 \\
&{\kern 50pt}+\frac{\lambda }{2}\left\| {{\mathbf{y}} - {\mathbf{Dx}}} \right\|_2^2,
\end{split}
\end{equation}
\[ s.t.{\kern 4pt} {{\bf{B}}_r} = {\cal R}{{\cal Q}_r}{\bf{U}},{\kern 5pt} {{\bf{C}}_r} = {\cal R}{{\cal Q}_r}{\bf{V}}.\]

In \eqref{eq:AuxiMode}, the first two terms are non-smooth but separable, and the other terms are smooth, which makes it relative easier than \eqref{eq:HVaF} in developing numerical algorithms. For a given $\beta $, we solve \eqref{eq:AuxiMode} by Alternating Direction Method of Multipliers (ADMM)\cite{Qu-Accelerate-2015}. The details of the new algorithm are omitted here due to limited space.

\section{Numerical experiments}
\label{sec:VI}

In this section, we will evaluate the performance of the proposed HVaF on synthetic data and real MRS. The state-of-the-art LRHMC \cite{Qu-Accelerate-2015} is compared with HVaF. The NUS is Poisson-gap sampling \cite{Hyberts2010}. We empirically observe that HVaF is very robust to $K$ and thus we set $K$ to be 64 in all the experiments of this paper.

\subsection{Synthetic data}
\begin{figure}[!ht]
\centering
\subfigure[]{
\begin{minipage}[t]{0.5\textwidth}
\centering
\includegraphics[width=2.8in]{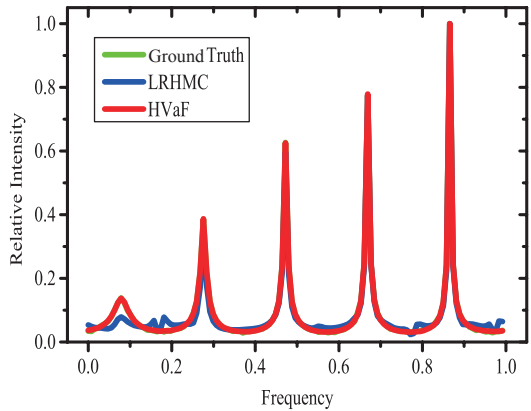}
\label{fig1a}
\end{minipage}
}
\vfill
\hspace{-3ex}
\subfigure[]{
\begin{minipage}[t]{0.5\textwidth}
\centering
\includegraphics[width=2.8in]{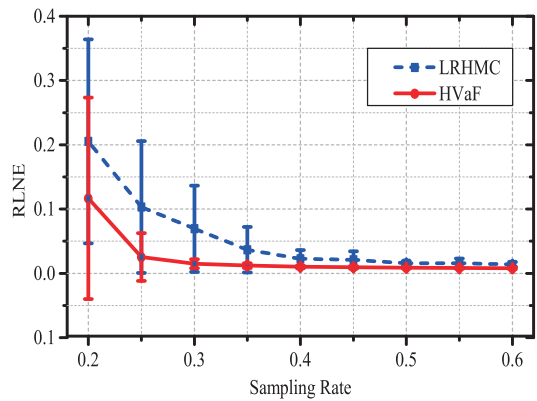}
\label{fig1b}
\end{minipage}
}
 \caption{ Reconstructions of the synthetic spectrum containing five peaks. Fig. \ref {fig1a} The reconstructions by LRHMC and HVaF from 25\% NUS. Fig. \ref{fig1b} The reconstruction RLNE with respect to different sampling rates.  The error bars are the standard deviations of the RLNE over 100 NUS resampling trials.}
\label{fig1}
\end{figure}

Fig. \ref{fig1} shows comparisons between a simulated fully sampled reference spectrum and its NUS reconstructions obtained using the LRHMC and HVaF algorithms. The simulated signals with 127 points are generated according to \eqref{eq:signal}. The relative least normalized error (RLNE) is defined by ${\left\| {{\mathbf{x}} - {\mathbf{y}}} \right\|_2}/{\left\| {\mathbf{y}} \right\|_2}$, where ${\bf x}$ and ${\bf y}$ are the reconstructed signal and true signal. It is observed that when sampling rate is relatively high, both LRHMC and HVaF can achieve a reconstruction with low RLNE; however, as sampling rate decreases, HVaF obtains a significantly lower RLNE than LRHMC. Thus HVaF requires fewer NUS samples to achieve faithful reconstructions than LRHMC. In particular, the line shape of the low-intensity peak can be preserved better by HVaF than LRHMC.

\subsection{Realistic MRS data}
\begin{figure*}[!ht]
\centering
\includegraphics[width=5in]{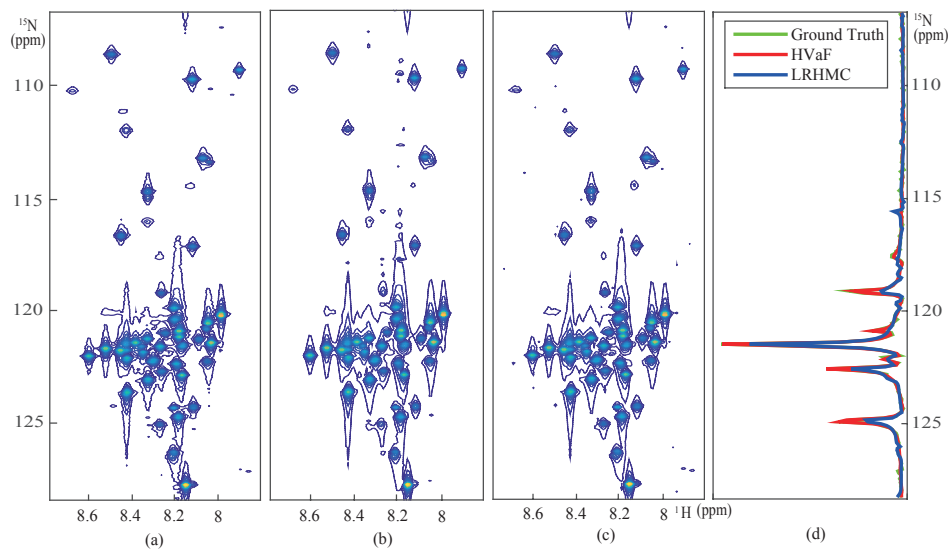}
\caption{The LRHMC and HVaF reconstructions in the 2D ${^1}$H-${^{15}}$N HSQC spectrum experiment. (a) the uniformly-sampled spectrum; (b) and (c) and are the LRHMC and HVaF reconstructions using 17\% sampled data, respectively. (d) the slice of the reconstructions located at the 8.25 ppm in the dimension of ${^1}$H.}
\label{fig2}
\end{figure*}

Here we apply the proposed HVaF to recover a 2-D spectrum with the size 512$\times$255 from the NUS FID. Fig \ref{fig2} shows a NUS 2D ${^1}$H-${^{15}}$N HSQC spectrum of the intrinsically disordered cytosolic domain of human CD79b protein from the B-cell receptor (More details about the MRS data and experiment setting can be found in \cite{Qu-Accelerate-2015}). Obviously, the HVaF produces the more faithful recovery (Fig. 2(c)) of the ground truth (Fig. 2(a)) than LRHMC (Fig. 2(b)). To clearly compare the reconstructed results, we present one of the slices in the reconstruction. Fig. 2(d) illustrates that  several low-intensity peaks are notably compromised in the LRHMC spectrum, while HVaF succeeds to recover most of the peaks.

\section{Conclusion and Discussion}
\label{sec:Discussion}
We develop the HVaF reconstruction as a new technique to obtain high-quality spectra from non-uniform sampling measurements. HVaF explores the Vandermonde structure of the Hankel matrix constructed by the FID instead of the low rank structure in LRHMC. Note that HVaF does not introduce extra assumptions on the signal compared with LRHMC. The method allows the more reduction in measurement time and achieves more faithful reconstruction than LRHMC. The Vandermonde factorization method will also be extended to the high-dimensional MRS reconstruction \cite{Ying2016}.

\section{Acknowledgments}
This work was supported by National Natural Science Foundation of China (61571380, 61302174 and U1632274), Fundamental Research Funds for the Central Universities (20720150109), Natural Science Foundation of Fujian Province of China (2015J01346, 2016J05205), Important Joint Research Project on Major Diseases of Xiamen City (3502Z20149032). The authors thank Vladislav and Maxim for sharing the MRS data used in paper \cite{Qu-Accelerate-2015}.

\bibliographystyle{IEEEtran}
\bibliography{MyBibliography}
\end{document}